\documentclass{elsart}
\usepackage{amssymb}
\usepackage{amsmath}
\begin{document}
\begin{frontmatter}
\title{An exact Lagrangian integral for the Newtonian gravitational field strength}
\author{Thomas Buchert}\ead{buchert@theorie.physik.uni-muenchen.de}
\address{Arnold Sommerfeld Center for Theoretical Physics, 
Ludwig--Maximilians--Universit\"at,
Theresienstr.~37, D--80333 M\"unchen, Germany}

\begin{abstract}
An exact expression for the gravitational field strength in a self--gravitating dust 
continuum is derived within the Lagrangian picture of continuum mechanics.
From the Euler--Newton system a transport equation for the gravitational field strength 
is formulated and then integrated along trajectories of continuum elements.
The resulting integral  solves one of the Lagrangian equations of the corresponding 
Lagrange--Newton system in general. Relations to known exact solutions without 
symmetry in Newtonian gravity are discussed. The presented integral may be employed 
to access the non--perturbative regime of structure formation in Newtonian cosmology,
 and to apply iterative Lagrangian schemes to solve the Lagrange--Newton system.
\end{abstract}

\end{frontmatter}

\section{The problem}
\label{sec:intro}

Let us characterize the state of a self--gravitating
``dust continuum'' (i.e., a continuum without pressure) at the initial time $t_i$ by
a velocity and a density field ${\bf v}({\bf x},t_i)$
and $\varrho ({\bf x},t_i)$. As usual, the fields are represented
in a non--rotating Eulerian coordinate system ${\bf x}$. We are interested in
the evolution of the continuum governed by the Eulerian evolution equations
for the velocity field ${\bf v}({\bf x},t)$ and the density field $\varrho({\bf x},t)$:
\begin{eqnarray}
\label{ENS1}
\frac{\partial}{\partial t}\Bigm\vert_{\bf x}
{\bf v} \;= - ({\bf v} \cdot \boldsymbol{\nabla}) {\bf v} + {\bf g} \;\; ;
\;\;{\bf v}({\bf x}, t_i) = :{\bf V} \;\;;\\
\frac{\partial}{\partial t}\Bigm\vert_{\bf x}
\varrho \;= - \boldsymbol{\nabla} \cdot(\varrho {\bf v}) \;\; ; \;\;
\varrho({\bf x}, t_i) = :\varrho_i  \;\; .\;\qquad
\label{ENS2}
\end{eqnarray}
The evolution of the continuum is constrained by Newton's
field equations for the gravitational field strength ${\bf g}({\bf x},t)$,
(which is equal to the acceleration field according to Einstein's
equivalence principle of inertial and gravitational mass):
\begin{eqnarray}
\boldsymbol{\nabla} \cdot {\bf g} = \Lambda - 4 \pi G \varrho \label{ENS3}\;\;;\\
\boldsymbol{\nabla} \times {\bf g} = {\bf 0} \;\;,\label{ENS4}
\end{eqnarray}
where $\Lambda$ is the cosmological constant, and the density field obeys
$\varrho \ge 0$. (The introduction of a gravitational potential
$\Phi$ defined by ${\bf g} =:- \boldsymbol{\nabla} \Phi$
(the existence of which is guaranteed by (\ref{ENS4})) will not be needed.)
Eqs.~(\ref{ENS1} -- \ref{ENS4}) form the {\it Euler--Newton system of equations}
\cite{heckmannschuecking:1955},
\cite{heckmannschuecking:1956}, \cite{heckmannschuecking:1959},
\cite{ehlers:1981}. 

Integration of this system is relevant in the context of Newtonian cosmology, where
the ``dust continuum'' models the (together with the source $\Lambda$)
dominant collisionless ``dark matter'' and ``dark energy''  in the Universe,
and is mostly studied perturbatively, i.e. perturbative solutions at a homogeneous--isotropic
solution of the Euler--Newton system are found in the Eulerian and Lagrangian
representations of the system (see, e.g., \cite{peebles:1980}, \cite{sahnicoles:1995}, 
\cite{ehlersbuchert:1997}, \cite{bernardeau:2002}
and references therein). Besides these analytical schemes the problem is commonly treated
by employing a variety of numerical integrators \cite{bertschinger:1998}.
Exact solutions for inhomogeneous fields 
are known for special spatial symmetries like
planar and spherical symmetry; three--dimensional solutions without symmetries  
are rare and will be put into perspective in Sect.\ref{sec:exactsolutions}.

\medskip

We proceed as follows.
We start by deriving a transport equation for the gravitational
field strength, working in the traditionally more emphasized Eulerian
picture (Sect.\ref{sec:transport}).
Then, we move to the Lagrangian picture  and integrate
the transport equation along the flow--lines of continuum elements, 
first by setting the cosmological
constant $\Lambda$ equal to zero, in Sect.\ref{sec:integral}. 
The result is generalized to the case of a 
non--vanishing cosmological constant in Sect.\ref{sec:lambda}. 
Sect.\ref{sec:exactsolutions}
provides relations to known exact solutions and presents an iteration scheme to solve
for the trajectory field.  Sect.\ref{sec:discussion} gives an
outlook on applications of the presented integrals in Newtonian cosmology.

\section{Transport equation for the gravitational field strength}
\label{sec:transport}

We first recall some equations that have been obtained earlier \cite{buchert:1989}, and
which provide an alternative formulation of the Euler--Newton system:
we can combine the evolution equation (\ref{ENS2}) with Eq.~(\ref{ENS3})
to obtain the following evolution equation for the gravitational field strength
(\cite{buchert:1989} Eq.~(7c*)):
\begin{equation}
\label{g-evolution-1}
\frac{d}{dt}  {\bf g} - \Lambda {\bf v} \;=\;
({\bf v} \cdot \boldsymbol{\nabla})
{\bf g} - {\bf v} (\boldsymbol{\nabla} \cdot {\bf g}) + \boldsymbol{\nabla} \times
\boldsymbol{\tau} \;\;,
\end{equation}
where $\frac{d}{dt}: = \frac{\partial}{\partial t}
\vert_{\bf x}+ {\bf v}
\cdot \boldsymbol{\nabla}$ is the total (Lagrangian) time--derivative,
abbreviated sometimes by an overdot.
(Hereafter, we shall drop the subscript $\vert_{\bf x}$ for notational ease.)

The vector field $\boldsymbol{\tau}$ specifies the freedom after
formally integrating the divergence of equation (\ref{g-evolution-1}), and can be
interpreted
as the vector potential of the current density ${\bf j} = \varrho {\bf v}$
(\cite{buchert:1989} Eq. (15a*)):
\begin{equation}
\label{g-evolution-2}
4 \pi G {\bf j} = - \boldsymbol{\nabla} \frac{\partial}{\partial t} \Phi - \boldsymbol{\nabla}
\times \boldsymbol{\tau} \;\;,
\end{equation}
where $\Phi$ is the gravitational potential as introduced above.
Eq.~(\ref{g-evolution-2}) is equivalent to Eq.~(\ref{g-evolution-1}) 
using the definition $\frac{d}{dt} {\bf g} = \frac{\partial}{\partial t} {\bf g} 
+ {\bf v} \cdot \boldsymbol{\nabla}{\bf g}$ and the source equation for the gravitational
field strength (\ref{ENS3}).

\medskip

In passing I mention that the field $\boldsymbol{\tau}$ plays the role of a
``magnetic'' field strength in a Maxwell--type analogy, while ${\bf g}$
corresponds to the ``electric'' field strength; we may cast Newton's field
equations together with Eq.~(\ref{g-evolution-2}) into the familiar form:
\begin{eqnarray}
&\boldsymbol{\nabla} \cdot {\bf g} = \Lambda - 4 \pi G \varrho\qquad
&\boldsymbol{\nabla} \cdot \boldsymbol{\tau} =  0 \;\;; \nonumber\\
&\boldsymbol{\nabla} \times {\bf g} = {\bf 0}\qquad\qquad\quad\,
&\boldsymbol{\nabla} \times \boldsymbol{\tau} = \frac{\partial}{\partial t} {\bf g}
- 4\pi G {\bf j} \;\;.
\end{eqnarray}
The constraint $\boldsymbol{\nabla} \cdot \boldsymbol{\tau} =  0$ is a gauge condition 
(transverse gauge) imposed on the vector potential $\boldsymbol{\tau}$. 
It is always possible to require this condition according to the
following reasoning:
the curl of the vector field $\boldsymbol{\tau}$ remains unchanged,
if we add to $\boldsymbol{\tau}$ the
gradient of an arbitrary scalar field, $\boldsymbol{\tau'}:=\boldsymbol{\tau}
+\boldsymbol{\nabla} \psi$; the divergence of
$\boldsymbol{\tau}$ itself is not specified by the Euler--Newton system, so we can
find $\psi$ such that $\boldsymbol{\nabla} \cdot \boldsymbol{\tau} = 0$ from the following
equation:
\begin{equation}
\label{gauge2}
\Delta \psi =  \boldsymbol{\nabla} \cdot \boldsymbol{\tau'} \;\;.
\end{equation}
There always exist solutions to Eq.~(\ref{gauge2})  according to a theorem
by Brelot (see, e.g., \cite{friedmann:1963}), if the spatial average of the source of 
Poisson's equation (\ref{gauge2}) vanishes.

Using this gauge condition,
we find from (\ref{g-evolution-2}) the simple relationship:
\begin{equation}
4\pi G \,\boldsymbol{\nabla} \times {\bf j} = -
\boldsymbol{\nabla} \times \left( \boldsymbol{\nabla} \times \boldsymbol{\tau} \right)
= \Delta \boldsymbol{\tau} - \boldsymbol{\nabla} \left(\boldsymbol{\nabla} \cdot 
\boldsymbol{\tau} \right) = \Delta \boldsymbol{\tau} \;\;.
\end{equation}

\medskip

Let us return to the derivation of the transport equation.
We employ the following vector identity:
\begin{equation}
\label{identity1}
\boldsymbol{\nabla}\times ({\bf g}\times {\bf v})\;=\;
({\bf v} \cdot \boldsymbol{\nabla}){\bf g} - {\bf v} (\boldsymbol{\nabla} \cdot {\bf g})
- \left[\, ({\bf g} \cdot \boldsymbol{\nabla}){\bf v}  -
{\bf g} (\boldsymbol{\nabla} \cdot {\bf v})\,\right] \;.
\end{equation}
With this identity, Eq.~(\ref{g-evolution-1}) can be cast into a transport
equation for the gravitational field strength:
computing $\varrho \frac{d}{dt}\left({\bf g}/\varrho\right)$, using (\ref{g-evolution-1}),
and inserting the continuity equation (\ref{ENS2}) in the form $\dot \varrho / \varrho
= -\boldsymbol{\nabla} \cdot {\bf v}$, we obtain:
\begin{equation}
\label{g-evolution-3}
\frac{d}{dt} \left(\frac{{\bf g}}{\varrho}\right) =
\left(\frac{{\bf g}}{\varrho} \cdot \boldsymbol{\nabla} \right) {\bf v}
+ \frac{\Lambda}{\varrho}{\bf v} + \frac{1}{\varrho} 
\boldsymbol{\nabla} \times \boldsymbol{\tilde\tau}\;\;\;;\;\;\;\boldsymbol{\tilde\tau}:=
\boldsymbol{\tau} + {\bf g} \times {\bf v}\;,
\end{equation}
which is the key--equation of the present work.
Eq.~(\ref{g-evolution-3}) (for $\Lambda=0$) is formally similar to 
Beltrami's transport equation for
the vorticity $\boldsymbol{\omega}: = \frac{1}{2}\boldsymbol{\nabla} \times {\bf v}$ 
reviewed in APPENDIX A.

\medskip

In the next section we shall integrate the transport equation 
(\ref{g-evolution-3}) after introducing
the Lagrangian form of the Euler--Newton system.

\section{Lagrangian integral of the transport equation for 
$\boldsymbol{\Lambda}{\bf = 0}$}
\label{sec:integral}

The Lagrangian description is based on integral curves
${\bf x} = {\bf f}({\bf X},t)$ of the velocity field
${\bf v}({\bf x},t)$:
\begin{equation}
\frac{d {\bf f}}{dt} = {\bf v} ({\bf f}, t) \;\; ; \;\;
{\bf f}({\bf X}, t_i) = :{\bf X} \;\; .
\end{equation}
In other words, $\bf f$ is the position vector field that locates a fluid element, indexed by 
its initial position vector (the Lagrangian coordinates $X_i$), in Eulerian space at a given
time $t$.
Introducing this family of trajectories, we can express
all Eulerian fields, e.g., the velocity ${\bf v}$, the gravitational
field strength ${\bf g}$, the density $\varrho$, and the vorticity $\boldsymbol{\omega}$
({\it cf.} APPENDIX A) in terms of the field of trajectories
${\bf x}={\bf f} ({\bf X},t)$ as follows:
\begin{eqnarray}
\qquad\qquad\qquad\qquad{\bf v} = \dot {{\bf f}} ({\bf X},t)\;\;;\;\;\label{lagint1}
{\bf g} = \ddot {{\bf f}} ({\bf X},t)\;\;; \\
\label{lagint2}
\qquad\qquad\qquad\qquad\varrho   = \varrho_i ({\bf X})\;J^{-1}\;\;;\\ \label{lagint3}
\qquad\qquad\qquad\qquad\boldsymbol{\omega} = \boldsymbol{\omega}_i
\cdot \boldsymbol{\nabla}_0 {{\bf f}} \;J^{-1}\;\;,
\end{eqnarray}
with the Jacobian of the transformation from Eulerian to Lagrangian coordinates
$J:= \det(f_{i | k}({\bf X},t))\;> 0$, 
$\boldsymbol{\omega}_i :=\boldsymbol{\omega} ({{\bf X}},t_i)$, and  
the nabla--operator with respect to Lagrangian coordinates denoted by $\boldsymbol{\nabla}_0$.
(Throughout this paper a vertical slash $|$ denotes partial differentiation with respect to 
Lagrangian coordinates, which commutes with the Lagrangian time--derivative, while
Eulerian spatial differentiation is abbreviated by a comma.)

Eqs.~(\ref{lagint2}) -- (\ref{lagint3}) are the known {\it Lagrangian integrals} 
of the Euler--Newton system, i.e., they represent an Eulerian field as a functional of $\bf f$.
As a result of the definitions (\ref{lagint1}) and the density integral (\ref{lagint2})
both Eulerian evolution equations are solved 
exactly {\it for any given trajectory field} in the Lagrangian picture.
Transforming those fields back to Eulerian space we need that the transformation
${\bf f}$ is invertible, i.e. $J>0$, defining {\it regular solutions} 
(for more details the reader may consult the review \cite{ehlersbuchert:1997}).

While the Eulerian evolution equations are represented by definitions and Lagrangian integrals,
the relevant equations in the Lagrangian picture are the Eulerian {\it field equations} 
(\ref{ENS3}) and (\ref{ENS4}), which are transformed into a system
of Lagrangian {\it evolution equations} by virtue of the following transformation of the 
field strength gradient $g_{i,j}= \frac{\partial}{\partial x_j}g_i$:
\begin{equation}
\label{g-gradient}
g_{i,j}=g_{i | k}h_{k,j} = \frac{1}{2J} \epsilon_{k\ell m}\epsilon_{jpq}
g_{i | k} f_{p | \ell} f_{q | m}\;\;;
\end{equation}
the gradient of the inverse transformation from Lagrangian to Eulerian coordinates,
${\bf h}={\bf f}^{-1}$ (which we require to exist), 
was expressed in terms of $\bf f$ through the algebraic relationship 
\begin{equation}
\label{inverseJacobian}
h_{i,j}  = J_{ij}^{-1} = ad(J_{ij})J^{-1} = \frac{1}{2J}\epsilon_{ik\ell}
\epsilon_{j m n} f_{m | k}f_{n | \ell}\;\;.
\end{equation}
In view of (\ref{ENS3}) and (\ref{ENS4}) we obtain with (\ref{g-gradient}) 
the following set of four Lagrangian equations
(\,\cite{buchertgoetz:1987}($\Lambda = 0$),
and \cite{buchert:1989} ($\Lambda \ne 0$); $i,j,k=1,2,3$ with cyclic
ordering; summation over repeated indices is understood):
\begin{eqnarray}
\label{lag1}
\qquad\qquad\frac{1}{2}\epsilon_{abc}\;\frac{\partial({g}_a,f_b,f_c)
}{\partial (X_1,X_2,X_3)} \; - \Lambda \,J
\; = \; - 4 \pi G \, \varrho_i ({\bf X})\; ;\\
\label{lag2}
\qquad\qquad\epsilon_{pq \lbrack j} \frac{\partial ({g}_{i\rbrack},
f_p,f_q)}{\partial(X_1,X_2,X_3)} = 0 \;\;\;,\; i \ne j \;\;.\qquad\;\;\qquad
\end{eqnarray}
Note that these equations only involve the dynamical variable ${\bf f}$
by virtue of ${\bf g} = {\ddot{\bf f}}$. Inserting this definition into the equations
above, we obain a set of four evolution equations furnishing 
the {\it Lagrange--Newton system}.
(Alternative forms of these equations may be found in \cite{buchert:1996} and
\cite{ehlersbuchert:1997}.)

\medskip

We now integrate the transport equation (\ref{g-evolution-3}) 
along trajectories ${\bf f}$ of continuum elements indexed by ${\bf X}$.
The  system (\ref{lag1},\ref{lag2}) is a set of nonlinear partial differential
equations coupled in a complicated way, and certainly more involved than their Eulerian
counterparts that are linear field equations. (Of course, the complete Eulerian system
is nonlinear due to the coupling of the field strength to the density, which itself has to 
obey a nonlinear equation.) With regard to the complicated nature of the Lagrangian 
equations it may be surprising that one can find -- as we shall see -- a relatively simple 
expression for $\bf g$ as a functional of $\bf f$, extending the set of known integrals 
Eqs.~(\ref{lagint2}) -- (\ref{lagint3}).
With the condition
\begin{equation}
\label{locality1}
\boldsymbol{\nabla} \times \boldsymbol{\tilde\tau} = {\bf 0}
\end{equation} 
we single out a class of motions that admits 
a {\it quasi--local} Lagrangian integral. The term `quasi--local' is
to be understood as follows: if we expect ${\bf g} = {\ddot{\bf f}}$ 
to be a functional of ${\bf f}$, then the trajectory
field is represented locally, while the initial data are still constructed
non--locally according to the structure of the theory, i.e. 
we have to prepare initial data such that all the field equations (\ref{ENS3},\ref{ENS4})
are satisfied initially.
However, we also 
expect a general integral, if it exists, to contain non--local parts that arise when 
the imposed restriction (\ref{locality1}) is relaxed. 
Requiring (\ref{locality1}) to hold is sufficient to enable us to
integrate Eq.~(\ref{g-evolution-3}), but it is not necessary for the following 
Proposition 1. 

Setting first $\Lambda=0$ we make the following ansatz:
\begin{equation}
\label{ansatz1}
\frac{{\bf g}}{\varrho} = (\bf s \cdot \boldsymbol{\nabla}_0) {\bf f} \;\;;\;\;
{\bf s} = {\bf s} ({\bf X},t) \;\;.
\end{equation}
Applying the total time derivative to this ansatz,
and using the identity
\begin{equation}
\label{identity2}
({\bf s} \cdot \boldsymbol{\nabla}_0) {\dot {{\bf f}}} = \left( ({\bf s} \cdot
\boldsymbol{\nabla}_0){\bf f} \right) \cdot \boldsymbol{\nabla} {\bf v} \;\;,
\end{equation}
we compare the resulting terms with
the transport equation (\ref{g-evolution-3}) to obtain:
\begin{equation}
({\dot {\bf s}} \cdot \boldsymbol{\nabla}_0) {\bf f} = {\bf 0}\;\;\;,\;{\rm i.e.}\;,\;{\rm in} \;\;
{\rm general}\;:\;\;
{\bf s} = {\rm const.} = \frac{\bf G}{\varrho_i}
\;\;;\;\;{\bf G} := {\bf g}({\bf x},t_i) \;\;.
\end{equation}
Using the integral for the density field, Eq.~(\ref{lagint2}), 
we arrive at an exact integral for the gravitational field strength
(this result has been first given in \cite{buchert:1993}, however, without providing the
derivation):

\bigskip

\noindent
{\bf Proposition 1} (Lagrangian Integral for $\Lambda = 0$)

\begin{itemize}
\item[]
The Lagrangian integral 
\begin{equation}
\label{integrala}
{{\bf g}}^{I}  =  \frac{({\bf G} \cdot \boldsymbol{\nabla}_0) {\bf f}}{J}\;\;\;;\;\;\;
J = \det (f_{i|k})
\end{equation}
solves the Lagrangian  equation (\ref{lag1}) for $\Lambda = 0$
exactly, while the remaining equations (\ref{lag2}) are reduced to constraint
equations restricting $\bf f$. 
\end{itemize}
(The proof follows by explicitly inserting the integral into the 
Lagrangian equations, either manually or using an algebraic
manipulation system, e.g. {\it REDUCE}. We outline an analytical proof in
APPENDIX B.)

\noindent
Note that this integral has its counterpart in the classical
integral (\ref{lagint3}) for the vorticity as reviewed in APPENDIX A.

\section{The case of a non--vanishing cosmological constant}
\label{sec:lambda}

The introduction of a cosmological constant in the Euler--Newton system 
corresponds to the introduction of a background field. In Newtonian cosmology
such a background is commonly installed in terms of a ``Hubble flow'', i.e., a  
homogeneous--isotropic solution of the Euler--Newton system that is furnished by 
the deformation field: 
\begin{equation}
\label{fhom}
{\bf f}_H = a(t){\bf X}\;\;\;\Rightarrow\;\;\;{\bf v}_H = \frac{\dot a}{a}{\bf x}\;\;;\;\;
\varrho_H = \varrho_H (t_i) a^{-3}\;\;,
\end{equation}
with the ``scale--factor'' $a(t)$, the linear ``Hubble velocity'' ${\bf v}_H$, 
and the homogeneous background density $\varrho_H (t)$.
Inserting ${\bf f}_H$ and ${\bf g}_H := {\ddot{\bf f}}_H$ 
into the Lagrange--Newton system (\ref{lag1}), (\ref{lag2}),
we obtain the well--known cosmological equation:
\begin{equation}
\label{friedmann1}
3\frac{\ddot a}{a} = \Lambda - 4\pi G \varrho_H\;\;.
\end{equation} 
Together with $\varrho_H$ as given in terms of $a(t)$ in (\ref{fhom}), this equation 
determines the solutions $a(t)$.
Deviation fields from the ``Hubble flow'' are then introduced. 
(Such a description lies at the basis of applications to cosmological
structure formation models, {\it cf.} Sect.\ref{sec:discussion}.)
In this line, the cosmological constant
plays the role of a static background, which introduces qualitatively new
features into the solutions.

Relaxing the restriction to a vanishing cosmological constant, 
we have to generalize the ansatz (\ref{ansatz1}): 
since $\Lambda$ is constant in space and
time, we expect a deviation from the field strength in (\ref{ansatz1}) that
is proportional to the displacement defined by ${\bf f}$:
\begin{equation}
\label{ansatz2}
\frac{{\bf g} - D{\bf f}}{\varrho} = ({\bf s} \cdot \boldsymbol{\nabla}_0) {\bf f} \;\;,
\end{equation}
with $D=const$.
Applying the total time derivative to the new ansatz (\ref{ansatz2}),
using again the identity (\ref{identity2})
and comparing the resulting terms
with the transport equation (\ref{g-evolution-3}) we first obtain:
\begin{equation}
\label{temp1}
\frac{D}{\varrho} \left[\; {\bf f} (\boldsymbol{\nabla} \cdot {\bf v})
- ({\bf f} \cdot \boldsymbol{\nabla}) {\bf v} +\frac{{\bf v}}{\varrho}\;\right]\;=\;
\Lambda \frac{{\bf v}}{\varrho}\;\;.
\end{equation}
With the following identity we are able to evaluate the constant $D$:
\begin{equation}
\label{identity3}
\boldsymbol{\nabla} \cdot \lbrack\;
{\bf f} (\boldsymbol{\nabla} \cdot {\bf v}) - ({\bf f} \cdot \boldsymbol{\nabla}) 
{\bf v} \;\rbrack =
\boldsymbol{\nabla} \cdot \lbrack\; {\bf v} (\boldsymbol{\nabla} \cdot {\bf f}) - 
({\bf v} \cdot \boldsymbol{\nabla}) {\bf f} \;\rbrack 
= \boldsymbol{\nabla} \cdot (d-1){\bf v} \;\;,
\end{equation}
where the dimension $d$ of the continuum has to be explicitly taken into
account. 

Since all equations are thus far understood up to a vector
potential that we neglected, we obtain in view of (\ref{temp1}) and
(\ref{identity3}): $D=\Lambda /d$. Using the Lagrangian integral for the density 
(\ref{lagint2}), we finally arrive
at a more general exact integral for the gravitational field strength:

\bigskip

\noindent
{\bf Proposition 2} (Lagrangian Integral for $\Lambda \ne 0$)

\begin{itemize}
\item[]
In $d$ dimensions of the continuum the Lagrangian integral
\begin{equation}
\label{integralb}
{{\bf g}}^{I\Lambda}  =  \frac{({\bf C} \cdot \boldsymbol{\nabla}_0) {\bf f}}{J}\;
+\frac{\Lambda}{d}\,{\bf f} \;\;;\;\;{\bf C}:={\bf G} - \frac{\Lambda}{d}\,{\bf X}\;\;;\;\;
J = \det (f_{i|k})
\end{equation}
solves the Lagrangian equation (\ref{lag1}) exactly, while the
remaining equations (\ref{lag2}) are reduced to constraint
equations restricting $\bf f$. 
\end{itemize}
(Again, the proof follows by explicitly inserting the integral into the 
Lagrangian equations, either manually or using an algebraic
manipulation system, e.g. {\it REDUCE}, see APPENDIX B for an analytical proof.)

\smallskip

\noindent
The explicit dependence of the integral (\ref{integralb}) on the dimension 
$d$ of the continuum can be easily understood by noting that from $\varrho =
0$, $\boldsymbol{\nabla} \cdot {\bf g} = \Lambda$, we have the integral ${\bf g} =
\frac{\Lambda}{d}{\bf f}$ which is confirmed by calculating the
divergence of this integral with respect to ${\bf x} = {\bf f} (\bf X,t)$.

\section{Relation to exact solutions and definition of an iteration scheme}
\label{sec:exactsolutions}

Let us first consider the integral (\ref{integrala}) for $\Lambda = 0$. 
It provides the (with respect to (\ref{lag1}) without any restriction) exact field strength along 
{\it any given} family of trajectories. Inserting the integral into the remaining
Lagrangian  equations (\ref{lag2}) delivers constraints that have to be
imposed on the given family of trajectories  yielding in particular restrictions on the
spatial coefficient functions.
(It is possible to find a corresponding exact integral for the Lagrangian  equations
corresponding to $\boldsymbol{\nabla}\times{\bf g}={\bf 0}$ which, again, requires
constraints by inserting it into (\ref{lag1}); this and also a formally general integral will 
be given elsewhere within a more general context; the integral of (\ref{lag1}) is favoured 
with regard to applications, {\it cf.} Sect.\ref{sec:discussion}.)

Alternatively, the integral (\ref{integrala})
may be regarded as a set of three partial differential equations for the components of the
trajectory field by virtue of the definition of ${\bf g}$ (\ref{lagint1}). 
Integrating the integral twice with respect to the time and fixing the two integration 
constants by the initial data: ${\bf f} ({\bf x},t_i)= {\bf X}$, ${\bf v} (\bf
x,t_i) = {\bf V}({\bf X})$, we obtain an integral equation for the trajectory field:
\begin{equation}
\label{iteration1}
{\bf f} = {\bf X} + {\bf V} ({\bf X}) (t-t_i) +
\int_{t_i}^t dt' \int_{t_i}^t dt' \;
({\bf G} ({\bf X})\cdot \boldsymbol{\nabla}_0) {\bf f} ({\bf X},t') J({\bf X},t')^{-1}
\;\;.
\end{equation}
This equation suggests, as
a manifest structural property of the integral (\ref{integrala}), 
the following {\it iterative definition} of the trajectory field ${\bf f}$:
\begin{equation}
\label{iteration2}
{{\bf f}}^{(n+1)} = {\bf X} + {{\bf V}} ({\bf X})(t-t_i) +
\int_{t_i}^t dt' \int_{t_i}^t dt' \;
({{\bf G}}({\bf X}) \cdot \boldsymbol{\nabla}_0) {{\bf f}}^{(n)}({\bf X},t')
\,J^{(n)}({\bf X},t')^{-1}\;\;,
\end{equation}
\begin{equation}
{\rm with}\qquad
J^{(n)}:=\det(f^{(n)}_{i|k})({\bf X},t')^{-1}\;\;,\nonumber
\end{equation}
where ${{\bf f}}^{(n)}$ is the $n^{th}$ iterate of the integral.
A trajectory field that is inserted on the r.--h.--s. of (\ref{iteration2})
does not in general yield the same trajectory field ${\bf f}$ on the
l.--h.--s. of (\ref{iteration2}).
This way we only consider ${\bf f}$ as a {\it solution} of the
Lagrangian equation (\ref{lag1}), iff ${\bf f}$ is,
in the language of dynamical systems theory,
a {\it fix--point} of this iteration, i.e.,
\begin{equation}
{{\bf f}}^{(n+1)} = {{\bf f}}^{(n)} \;\;,
\end{equation}
for all $n$ larger than some $n^*$.

Let us now consider some known cases of exact solutions in Newtonian
gravity.
Among the simplest of exact solutions we have already introduced the
homogeneous--isotropic deformation fields ${{\bf f}}_H$ (\ref{fhom}) and
(\ref{friedmann1}).
Inserting this solution for $\Lambda = 0$
into the r.--h.--s. of (\ref{iteration2}),
we recover the same solution on the l.--h.--s. of (\ref{iteration2})
(with careful handling of the integration constants).
Thus, in this case, the iterated solution is identical to the solution
fed into the integral.

\noindent
We face
the same situation in the case of the exact general one--dimensional solution
\cite{novikov:1970}, \cite{buchertgoetz:1987}:
\begin{equation}
f_1 = X_1 + V_1 (X_1) (t-t_i) + G_1 (X_1)\frac{(t-t_i)^2}{2} \;\;.
\end{equation}
Also, the three--dimensional generalization of this solution \cite{buchertgoetz:1987},
admitting locally (at each $\bf X$ different) one--dimensional motions and imposing
no global symmetry restrictions, forms a {\it fix--point} of this iteration:
\begin{equation}
\label{3Dsolution}
{\bf f} = {\bf X} + {\bf V} ({\bf X})(t-t_i) + {\bf G}({\bf X}) \frac{(t-t_i)^2}{2}\;\;.
\end{equation}
Here, we have to make explicit use of the constraints
given in \cite{buchertgoetz:1987} and the identity (\ref{identity1}) to verify this.
(The solution class is defined by three--dimensional initial data that are composed of 
2--surfaces with vanishing Gaussian curvature; the plane--symmetric case is contained
in a subclass formed by cylindrical 2--surfaces.)
The basic assumption that was used to derive this class of exact
solutions was the constancy of the gravitational field strength
along the solution curves: ${\bf g} ({\bf X},t)= {\bf G} ({\bf X})$. The
integral (\ref{integrala})
can be viewed as a generalization of this assumption, where ${\bf g}$
changes along solution curves according to their directional derivative
with respect to the initial field ${\bf G}$, weighted in addition by the local density.
It is interesting that we obtain the solution curves (\ref{3Dsolution})
as the first iterate of the integral (\ref{integrala}), if we start
with the trivial trajectories ${{\bf f}}^{(0)} = {\bf X}$.
Any further iterate is identical to ${{\bf f}}^{(1)}$, if the constraints
quoted above are respected.

Of course, we may start this iteration scheme with generic initial data in order to
approximate solutions of the Lagrangian system.
Iteration will produce another approximation that will contain,
at each iteration step, higher spatial derivatives of the initial data. 
At this stage it is premature to expect that such an iteration scheme may converge in the sense of producing 
more refined models which approach a solution of the 
Lagrange--Newton system. 

\medskip

Relaxing the restriction $\Lambda = 0$, we accordingly obtain the more
general trajectory field, which we may write as follows:
\begin{equation}
\label{3DsolutionLambda}
{\bf f} = {\bf X} + {\bf V}({\bf X}) (t-t_i)+
\int_{t_i}^t dt' \int_{t_i}^t dt'\;
\left[\,({\bf G}({\bf X}) - \frac{\Lambda}{d} {\bf X}) \cdot 
\boldsymbol{\nabla}_0 {\bf f}\, J^{-1} +\frac{\Lambda}{d}\,{\bf f}\,\right] \;\;.
\end{equation}
Viewing the above equation as an iteration scheme, then
one possible choice is to write:
\begin{equation}
\label{iterationchoice1}
{\ddot {{\bf f}}}^{(n+1)} - \frac{\Lambda}{d} {{\bf f}}^{(n+1)} = \frac{({\bf G}({\bf X}) -
\frac{\Lambda}{d}\,{\bf X}) \cdot \boldsymbol{\nabla}_0 
{{\bf f}}^{(n)}}{J({{\bf f}}^{(n)})}\;\;.
\end{equation}
If we start with a homogeneous--isotropic deformation, 
${{\bf f}}^{(0)} = a (t) {\bf X}$, then we
find a {\it fix--point} corresponding to all solutions of Eqs.~(\ref{fhom}), 
(\ref{friedmann1}) for $d=3$. 

On the other hand, if we start with the trivial trajectories ${{\bf f}}^{(0)}={\bf X}$,
then the first iterate obeys:
\begin{equation}
{\ddot {{\bf f}}}^{(1)} - \frac{\Lambda}{d} {{\bf f}}^{(1)} = {\bf G}({\bf X}) -
\frac{\Lambda}{d}\,{\bf X} \;\;,
\end{equation}
admitting the general integral--curves:
\begin{eqnarray}
\label{generalLambda1}
&{{\bf f}}^{(1)} = {\bf X} + \frac{d}{2\Lambda}
\left({\bf G}({\bf X}) +\sqrt{\frac{\Lambda}{d}}\;
{\bf V} ({\bf X})\right) \left[ e^{\sqrt{\frac{\Lambda}{d}}\;
(t-t_i)}-1 \right] \nonumber\\
&\qquad\qquad\quad+\frac{d}{2\Lambda} \left(
{\bf G}({\bf X}) - \sqrt{\frac{\Lambda}{d}}\; {\bf V}({\bf X}) \right)
\left[ e^{- \sqrt{\frac{\Lambda}{d}}(t-t_i)}-1 \right] \;\;.
\end{eqnarray}
For the case $d=3$ we propose a more refined choice of the iteration scheme 
(\ref{iterationchoice1}), motivated by the idea that the first iterate then provides an
exact solution to the Lagrange--Newton system (\ref{lag1}), (\ref{lag2}). We write:
\begin{equation}
\label{iterationchoice2}
{\ddot {{\bf f}}}^{(n+1)} - \Lambda {{\bf f}}^{(n+1)} = \frac{({\bf G}({\bf X}) -
\frac{\Lambda}{3}\,{\bf X}) \cdot \boldsymbol{\nabla}_0 {{\bf f}}^{(n)}}
{J({{\bf f}}^{(n)})}\;-\;\frac{2\Lambda}{3}{{\bf f}}^{(n)}\;\;.
\end{equation}
With this choice the Friedmannian solutions still form a fix--point 
corresponding to the iteration of a homogeneous--isotropic deformation (as above). 
The advantage of this
latter choice becomes obvious by iteration of the trivial trajectories 
${{\bf f}}^{(0)}={\bf X}$, since the first iterate then obeys:
\begin{equation}
{\ddot {{\bf f}}}^{(1)} - \Lambda {{\bf f}}^{(1)} = {\bf G}({\bf X}) -\Lambda\,{\bf X} \;\;,
\end{equation}
admitting the general integral--curves:
\begin{eqnarray}
\label{generalLambda2}
&{{\bf f}}^{(1)} = {\bf X} + \frac{1}{2\Lambda}
\left({\bf G}({\bf X}) +\sqrt{\Lambda}\;
{\bf V} ({\bf X})\right) \left[ e^{\sqrt{\Lambda}\;
(t-t_i)}-1 \right] \nonumber\\
&\qquad\qquad\quad+\frac{1}{2\Lambda} \left(
{\bf G}({\bf X}) - \sqrt{\Lambda}\; {\bf V}({\bf X}) \right)
\left[ e^{- \sqrt{\Lambda}(t-t_i)}-1 \right] \;\;.
\end{eqnarray}
The trajectory field (\ref{generalLambda2}) provides a class of three--dimensional
exact solutions without global symmetry restrictions that is contained in a subclass
of solutions investigated in \cite{buchert:1989}, which will not be further discussed here
(see also \cite{bildhauerbuchert:1991}; for solutions including a cosmological constant
see \cite{bildhauerbuchertkasai:1992}, \cite{cherninetal:2003}, 
and for related rotational solutions \cite{silbergleit:1995}).
Explicitly, this solution class has been
discussed by Barrow \& G\"otz \cite{barrowgoetz:1989}
in the context of ``no--hair'' theorems.

Although the basic equations tend continuously, in the limit $\Lambda \rightarrow 0$,
to their restricted set for $\Lambda = 0$,
the solution (\ref{3Dsolution}) given above can only be obtained from 
(\ref{generalLambda2})
as an {\it approximate} limit by expanding the exponentials.
This shows that the presence of a background entails a qualitative difference.

\section{Applications in Newtonian cosmology}
\label{sec:discussion}

In order to appreciate the application side of the presented integrals, let us give a 
concrete example in the context of structure formation models in Newtonian cosmology,
which requires the presence of a background.

We employ the integral (\ref{integralb}) for $d=3$. 
Since this integral is exact, we are entitled to apply 
the transformation to a ``comoving trajectory field'', ${\bf q}=:{\bf F}({\bf X},t)=
{\bf x}/a(t)$, and to introduce deviation--fields from the background solution:
since we can adopt the same Lagrangian coordinates $X_i$ in both cases in view of 
${\bf x}= a(t){\bf q}$, $a(t_i) = 1$, ${\bf F}({\bf X},t_i) = {\bf X}$, and since the
background field strength is evolving as ${\bf g}_H = {\ddot a}{\bf q}$, we make the
ansatz ${\bf g}= {\ddot a}{\bf F} + {\bf w}({\bf X},t)$,  
${\bf G}= {\ddot a}(t_i){\bf X} + {\bf W}({\bf X})$, use Friedmann's equation 
(\ref{friedmann1}), and obtain from (\ref{integralb}) the following transformed 
integral for the ``peculiar--gravitational field strength'' ${\bf w}$ in terms of $\bf F$ :
\begin{equation}
\label{w_exactintegral}
{\bf w}^{I} \; = \; 
\frac{\left(\,{\bf C} \cdot \nabla_{0}\,\right) {\bf F}}{a^2 J_F} \;+ \;
\frac{4 \pi G\varrho_H a}{3}\,{\bf F} \;\;;\;\;J_F :=\det(F_{i|k})=J a^{-3}\;\;;
\end{equation}
${\bf C} = {\bf W} - \frac{4 \pi G \varrho_{H}(t_i)}{3}{\bf X}$ is the integration constant.
The iteration scheme is defined, for $d=3$, according to the choice 
(\ref{iterationchoice2}):
\begin{equation}
\label{w_exactiteration}
\frac{1}{a}{\bf w}^{[n+1]} - 4 \pi G\varrho_H {\bf F}^{[n+1]}  =  
\frac{\left(\,{\bf C} \cdot \nabla_{0}\,\right) {\bf F}^{[n]}}{a^3 J_F^{[n]} } - 
\frac{8 \pi G\varrho_H}{3}{\bf F}^{[n]} \;,
\end{equation}
with $\frac{1}{a}{\bf w}^{[n+1]} 
= {\ddot{\bf F}}^{[n+1]} + 2H {\dot{\bf F}}^{[n+1]}$. 
The idea is to start this iteration scheme for generic
initial data defining an {\it approximation} to the Lagrange--Newton system.
Starting the iteration (\ref{w_exactiteration}) with the trivial trajectory field 
${\bf F}^{[0]}= {\bf X}$, which 
corresponds to a ``Hubble flow'', we obtain the equation governing 
the longitudinal first--order Lagrangian perturbation solution in Newtonian cosmology
\cite{buchert:1992}, ${\bf F}^{[1]}= {\bf X}+{\bf P}^{[1]}$, with ${\bf P}^{[1]}$ given
for all Friedmannian backgrounds in \cite{bildhauerbuchertkasai:1992};
a subclass of this trajectory field is the celebrated
``Zel'dovich approximation'' \cite{zeldovich:1970}, which is already very successful 
for the description of structure formation in cosmology in comparison with N--body 
simulations (see the references in the review papers \cite{bouchetetal:1995},  
\cite{sahnicoles:1995}, \cite{ehlersbuchert:1997}). Any further
iterate produces a non--perturbative approximation. A numerical implementation of 
this iteration scheme  is currently under investigation.  

\smallskip

A final word on the range of validity of the presented integrals is in order. Although we
discussed all equations and results in the framework of a ``dust continuum'', 
Equations (\ref{g-evolution-3}),
(\ref{lag1}), (\ref{lag2}) and their integrals for ${\bf g}$ are valid for a much wider range of
continua, e.g. including pressure forces. In those more general cases the representation
of  ${\bf g}$ in terms of the trajectory field changes, and is no longer given by the simple
relationship   ${\bf g}=\ddot {\bf f}$ as in the ``dust continuum'' (see, e.g., 
\cite{adlerbuchert:1999} for the case of an isotropic pressure).
This more general context and all details relevant to applications in Newtonian cosmology
will be presented in a forthcoming paper  \cite{buchert:2006}.


\subsection*{Acknowledgements}

This work was supported by the `Sonderforschungsbereich SFB 375 
Astroparticle physics' by the German science foundation DFG.
Thanks go to Uriel Frisch, Andrei Sobolevskii and the referee for valuable comments.


\subsection*{APPENDIX A: Lagrangian integral for the vorticity}

\vspace{-3pt}

We here review a classical result that displays formal similarities to the
integration procedures used for the derivation of the transport equation for the gravitational
field strength. 
The transport equation (\ref{g-evolution-3}) is formally similar to 
Beltrami's transport equation for the vorticity 
$\boldsymbol{\omega} : = \frac{1}{2}\boldsymbol{\nabla}
\times {\bf v}$ (see, e.g., \cite{serrin:1959} and \cite{buchert:1992}):
\begin{equation}
\frac{d}{dt} \left(\frac{\boldsymbol{\omega}}{\varrho}\right) \;=\; 
\left(\frac{\boldsymbol{\omega}}{\varrho} \cdot \boldsymbol{\nabla} \right){\bf v} +
\frac{1}{\varrho} \boldsymbol{\nabla} \times {\bf g} \;\;.
\end{equation}
For curl--free forces, which is the case for Newtonian flows,
this reduces to:
\begin{equation}
\frac{d}{dt} \left(\frac{\boldsymbol{\omega}}{\varrho}\right) \;=\; 
\left(\frac{\boldsymbol{\omega}}{\varrho} \cdot \boldsymbol{\nabla} \right){\bf v} \;\;.
\end{equation}
There is also a classical integral of Beltrami's
transport equation due to Cauchy in the case of curl--free
forces (see \cite{serrin:1959} and \cite{buchert:1992}):
\begin{equation}
\boldsymbol{\omega} =  \frac{\boldsymbol{\omega}_i
\cdot \boldsymbol{\nabla}_0 {{\bf f}}}{J}\;\;;\;\;
\boldsymbol{\omega}_i :=\boldsymbol{\omega} ({{\bf X}},t_i)\;\;.
\end{equation}

\subsection*{APPENDIX B: Proof of Propositions 1 and 2}

\vspace{-3pt}

A simple way to prove Proposition 2 and, being a subcase, Proposition 1
is to write the divergence of ${\bf g}$ as $g_{i,i} = g_{i|k}h_{k,i}$ (first without explicitly
writing out ${\bf h}$).
Computing $ g_{i|k}$ from the integral (\ref{integralb}), 
$g_i = C_{s}f_{i|s} J^{-1} + \frac{\Lambda}{d} \, f_i$, we first have:
\begin{equation}
g_{i|k} = \frac{1}{J}\left[\,C_{s|k}f_{i|s} + C_s f_{i|sk} - \frac{J_{|k}}{J}C_s f_{i|s}
\,\right] + \frac{\Lambda}{d}\,f_{i|k}\;\;.
\end{equation}
Multiplying by $h_{k,i}$ and using $f_{i|s}h_{k,i}=\delta_{sk}$, we obtain:
\begin{equation}
\label{appB1}
g_{i,i} =  \frac{1}{J}\left[\,C_{k|k} + C_s f_{i|sk}h_{k,i} - \frac{J_{|k}}{J}C_k
\,\right] + \frac{\Lambda}{d}\,\delta_{kk}\;\;.
\end{equation}
We have to prove that $g_{i,i}=\Lambda - 4\pi G \varrho_i  J^{-1}$. Since
$C_k = G_k - \frac{\Lambda}{d}\,X_k$, $G_{k|k} = \Lambda-4\pi G \varrho_i$,
$X_{k|k}=\delta_{kk}=d$,
the first term in Eq.~(\ref{appB1}) together with $\frac{\Lambda}{d}\,\delta_{kk}
=\Lambda$, 
already provides the whole equation, $g_{i,i}=\Lambda - 4\pi G \varrho$.
It remains to prove that, for $J\ne 0$, the other two terms in the brackets of
Eq.~(\ref{appB1}) cancel out:
\begin{equation}
C_s f_{i|sk}h_{k,i} = \frac{J_{|k}}{J}C_k\;\;.
\end{equation}
We now write the inverse matrix explicitly, 
$h_{s,i} = \frac{1}{2J} \epsilon_{ipq}\epsilon_{sro}\,f_{p|r}f_{q|o}$, 
and also the Jacobian, $J= \frac{1}{6}   \epsilon_{abc}\epsilon_{def}\,
f_{a|d}f_{b|e}f_{c|f}$ to obtain the equivalent requirement:
\begin{equation}
C_s \epsilon_{ipq}\epsilon_{sro}\,f_{i|sk}f_{p|r}f_{q|o} = C_k
 \frac{1}{3}   \epsilon_{abc}\epsilon_{def}\left(\,f_{a|d}f_{b|e}f_{c|f}\,\right)_{|k}\;.
\end{equation}
We appreciate that, by evaluating the r.--h.--s., we get three identical expressions, so that by 
relabelling the summation indices we conclude that the above equation 
holds.$\;\blacksquare$

\vspace{10pt}

\noindent
{\sl The REDUCE codes to compute the Lagrangian evolution equations and the
presented integrals can be obtained from the author.}

\end{document}